\documentclass[a4paper,aps,prd,twocolumn,showpacs,eqsecnum,amsmath,amssymb,nofootinbib,superscriptaddress,float]{revtex4-1}
\usepackage{graphicx}
\usepackage{multirow}
\usepackage[table,xcdraw]{xcolor}
\usepackage{hhline}
\usepackage{booktabs}
\usepackage{slashed}
\usepackage{rccol}
\usepackage{mathrsfs}
\usepackage{dcolumn}
\usepackage{bm}

\usepackage{xcolor}\colorlet{linkequation}{blue}
\usepackage[colorlinks=true, 
            linkcolor=red,   
            filecolor=red,   
            citecolor=green, 
            urlcolor=blue,
            hyperfootnotes=true]{hyperref}

\newcommand*{\SavedEqref}{}
\let\SavedEqref\eqref
\renewcommand*{\eqref}[1]{%
  \begingroup
    \hypersetup{
     linkcolor=linkequation,
      linkbordercolor=linkequation,
    }%
    \SavedEqref{#1}%
  \endgroup
}

\newcommand{\be}{\begin{equation}}
\newcommand{\ee}{\end{equation}}
\newcommand{\bq}{\begin{eqnarray}}
\newcommand{\eq}{\end{eqnarray}}

\begin{document}

\title{Accelerating Shor's Factorization Algorithm on GPUs}
\author{I.~Savran}
\email{savran@ktu.edu.tr}
\affiliation{Department of Computer Engineering and}%
\author{M.~Demirci}
\email{mehmetdemirci@ktu.edu.tr}
\author{A. H.~Y{\i}lmaz}
\email{hakany@ktu.edu.tr}
\affiliation{Department of Physics, Karadeniz Technical University, 61080 Trabzon, Turkey}%

\date{\today}

\begin{abstract}
Shor's quantum algorithm is very important for cryptography, since it can factor large
numbers much faster than classical algorithms.
In this study, we implement a simulator for Shor's quantum algorithm on graphic processor units (GPU) and
compare our results with Liquid -which is Microsoft quantum simulation platform- and two classical CPU-implementations.
We evaluate 10 benchmarks for comparing our GPU implementation with Liquid and single-core implementation.
The analysis shows that GPU vector operations is more suitable for Shor's quantum algorithm. Our GPU kernel function is compute-bound, due to all threads in a block reach to the same element of the state vector. Our implementation has $52.5\times$ speedup over single-core algorithm and $20.5\times$ speedup over Liquid.
\end{abstract}

\maketitle

\section{\label{sec:level1} Introduction}
In the past thirty years, quantum computing has received considerable attention regarding performance speedup in the IT Society. It focuses on developing computer technology based on the principles of quantum theory such as superposition, entanglement and interference. Especially, it can accept input states that represent a coherent superposition of many different possible inputs and then turn them into a corresponding superposition of outputs. Quantum computation have an effect simultaneously on each element of the superposition and it generates a massive parallel data processing, even within one piece of quantum hardware. It enables large improvements in computational efficiency such as Shor's quantum algorithm for factoring large integers~\cite{shor1,shor2}, Grover's algorithm for accelerating combinatorial searches~\cite{Grover} and quantum cryptography for secure communication~\cite{crypt,crypt2}. As a result, some problems that are difficult to solve in a classical computer can be effectively solved by a quantum computer. One problem of this type is factorization.

A well-known quantum algorithm for factorization is Shor's algorithm. The computational resources increase exponentially with system size, and the simulation results are easily confirmed, making it an ideal test candidate. Many cryptographic protocols are based on the computational difficulty of obtaining the prime factors of a large number: a small increase in the size of the number causes an exponential increase in computational resources. However, such limitation does not appear in Shor's quantum algorithm for prime number factorisation, and its realization represents a major challenge in quantum computation. This algorithm is a set of protocols that convert the factorization problem into a period detection problem.

Since the announcement of the NVIDIA Compute Unified Device Architecture (CUDA)~\cite{NVIDIA} in 2008, graphic processor units (GPUs) computing has become widely adopted by the computing society. The GPU executes one or more kernels launched by the CPU. As compared to CPUs, a larger portion of their resources are devoted to functional units in GPUs. GPUs use smaller, nonexpandable DRAMs but have substantially higher memory bandwidth than CPUs. Most of the large-scale supercomputer installations equipped with computing accelerators. Intel Xeon Phi, NVIDIA and AMD Radeon are well known computing accelarators.

In this work we implement a simulator for Shor's quantum algorithm on GPUs in order to compute the prime factors of a few integers. The inherit parallelism involved in simulating a quantum system makes it suitable for on GPUs implementations. We compare our results with those obtained from Liquid-Microsoft quantum simulation platform. In ref~\cite{Wang2017}, they have reported results applying matrix product state to Shor's algorithm. 

The remainder of this work is organized as follows. Section~\ref{sec:shor} provides a review of Shor's algorithm.
Section~\ref{sec:implementation} describes our implementation in
detail. Section~\ref{sec:results} provides the numerical results related to finding the prime factors of several number and discusses for different implementations. Section~\ref{sec:conc}  summarizes the highlights
of the study.

\section{\label{sec:shor} A Brief Introduction to Shor's algorithm}
We describe Shor's algorithm from a functional point of view which means that it doesn't deal with the implementation for a specific hardware architecture. A detailed information and implementation about the Shor's algorithm can be found in references \cite{shor3, compt,compt2}, for a more rigid mathematical description, please refer to \cite{compt3, Hayward}.

Shor's algorithm tries to find even integer $p$, the period of $x^a~\text{mod}~n$, where $n$ is the number to be factored and $x$ is an integer coprime to $n$. To do this, a quantum memory register with two parts is created by Shor's algorithm as follows: $|r_1,r_2\rangle$. In the first part a superposition of the integers which are to be $a$'s in the $x^a~mod~n$ function is placed by the algorithm. Here, $a$'s can be chosen to be the integers $0$ through $q - 1$, where $q$ is the power of $2$ such that $n^2 \leq q < 2n^2$. In this step, the state of the quantum memory register is
\begin{equation}\label{qmr}
\frac{1}{\sqrt{q}}\sum^{q-1}_{a=0}|a,0\rangle.
\end{equation}
Then $x^a~mod~n$ is calculated, and the result is placed in the second part of the quantum memory register.

Next the state of the second register is measured by the algorithm, the one that includes the superposition of all possible outcomes for $x^a mod~n$. In this step, the state of the quantum memory register is given by
\begin{equation}\label{qmr}
\frac{1}{\sqrt{q}}\sum^{q-1}_{a=0}|a, x^a mod~n\rangle.
\end{equation}
Measuring this register has the effect of collapsing the state into some observed value, say $k$. This means that after this measurement the second part of the register contains the value $k$, and the first part of the register contains a superposition of the base states which when plugged into $x^a~mod~n$ produce $k$. Because $x^a~mod~n$ is a periodic function, the first part of the register will contain the values $c, c + p, c + 2p, \ldots $ and so on, where c is the lowest integer such that $x^c~mod~n=k$.

The next step is to perform a discrete Fourier transform (DFT) on the contents of first part of the register and to put the result back into register one. DFT when applied to a state $|a\rangle$ changes it in the following manner:
\begin{equation}\label{dft}
 |a\rangle=\frac{1}{\sqrt{q}}\sum^{q-1}_{c=0}e^{i2 \pi a c/q}|c\rangle
\end{equation}
which is computed by the quantum computer. The application of the DFT has the effect of peaking the probability amplitudes of the first part of the register at integer multiples of the quantity $q/p$.
Now measuring the first part of the quantum register will give an integer multiple of the inverse period. Once this number is retrieved from the quantum memory register, a classical computer can do some analysis of this number, make an estimation as to the actual value of $p$, and from that compute the possible factors of $n$.

\section{\label{sec:implementation} Implementation }
We now describe the details of our simulations and present our benchmarks. In conventional implementation, approximately 97\% of the runtime is devoted to Quantum Fourier Transform (QFT) calculations. Therefore, we designed a GPU kernel function which computes QFT from scratch. The computations related to the evolution of the quantum system are carried out by thousands of threads inside a GPU. Each thread is assigned to compute each element of the result vector. This approach allows threads to access memory coalescingly.

\subsection*{\label{sec:QFT} The Quantum Fourier Transform}
In quantum computing, QFT is a linear transformation on qubits and an essential part of many quantum algorithms, such as Shor's factoring algorithm. The quantum computers can perform QFT efficiently, with a particular decomposition into a product of simpler unitary matrices. Using a simple decomposition, DFT can be implemented as a quantum circuit consisting of only $O(n^2)$ Hadamard and controlled phase shift gates, where $n$ is the number of qubits.

\begin{figure}[hbpt]
    \begin{center}
\includegraphics[scale=0.35]{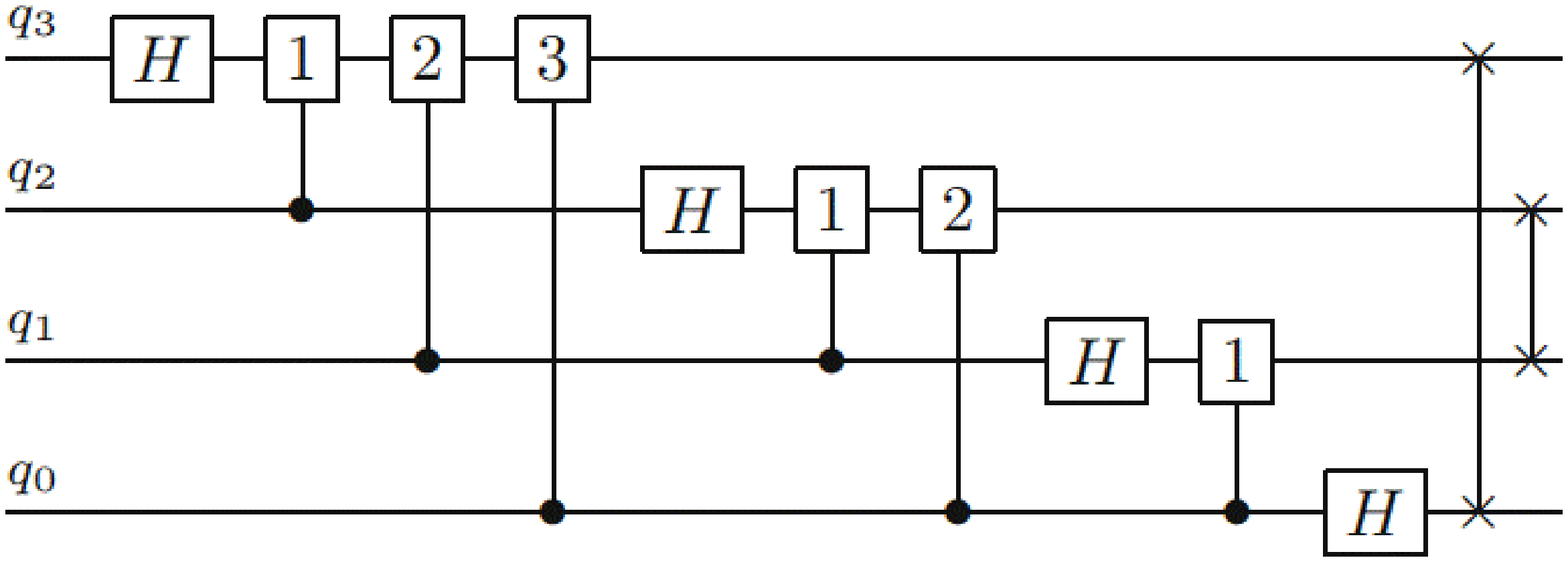}
     \end{center}
\caption{Standard quantum circuit for QFT on four qubits. $H$ is the Hadamard gate and other operators are controlled-phase gates.} \label{Fig1}
\end{figure}

A typical four-qubit quantum circuit for the QFT is shown in Figure~\ref{Fig1}.

\begin{figure*}[t]
      \includegraphics[scale=0.50]{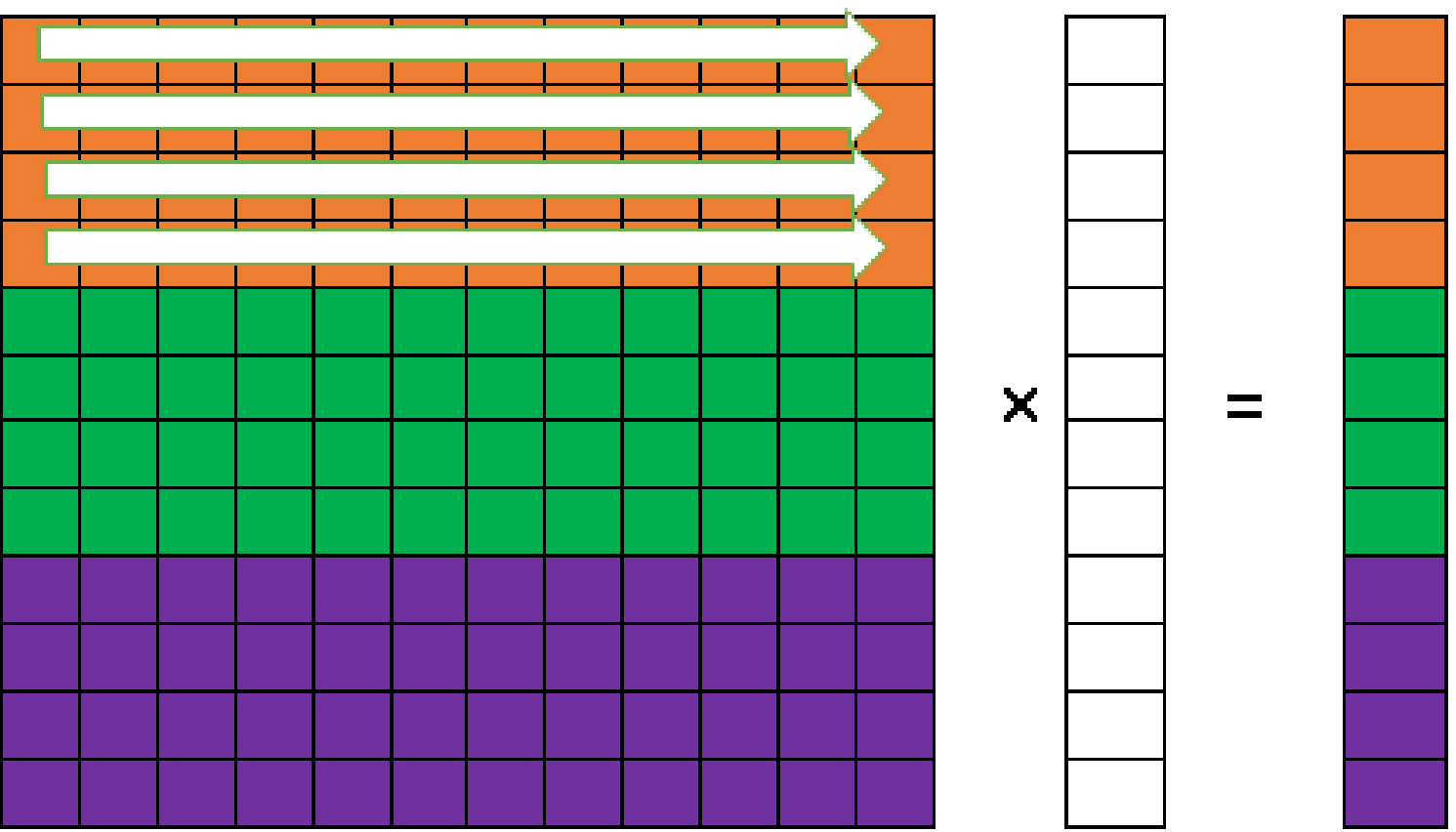}\hfill
      \includegraphics[scale=0.50]{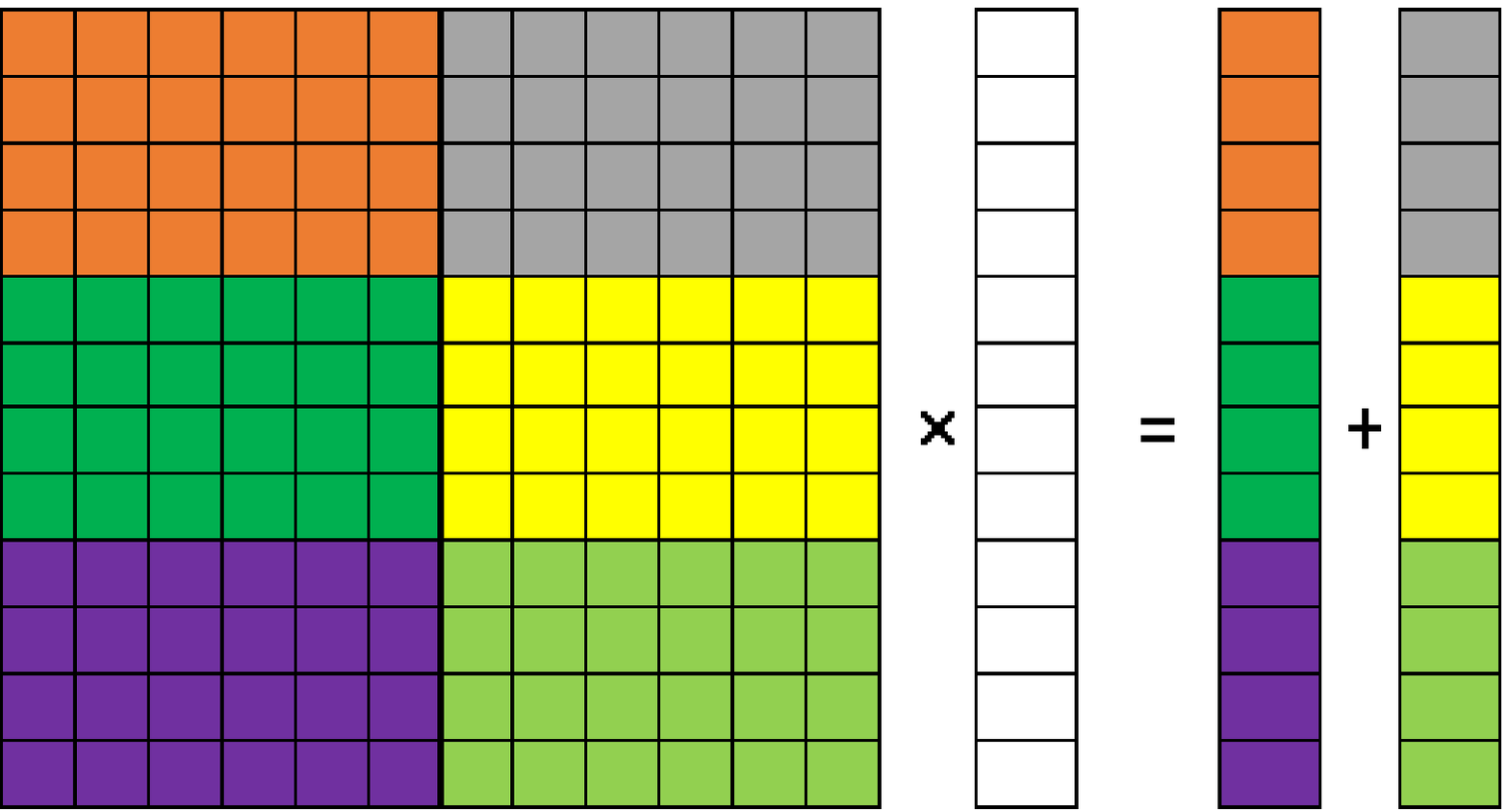}
  \caption{(color online). (a) Block configuration and threads running order in GPU. Each color indicates a different block of threads. (b) By using more blocks, the workload of the threads can be reduced.}\label{Fig2}
\end{figure*}
\begin{table*}[t]
\caption{Comparison of CPU with GPUs used in our tests.}\label{table1}
\begin{ruledtabular}
\begin{tabular}{lccc}
 &\multicolumn{1}{c}{\textbf{Intel I7-2760QM}}&\multicolumn{1}{c}{\textbf{NVIDIA GTX285}}&\multicolumn{1}{c}{\textbf{NVIDIA 970m}}\\ \hline
Architecture&Sandy Bridge&Tesla 2.0&Maxwell\\
Processor Core&4&30 (8 shaders)&10 (128 shaders)\\
Threads/Core&2&512&2048\\
Clock Frequency&2.2 GHz&648 Mhz&924 MHz\\
Memory Bandwidth&21.3 GB/s&159 GB/s&120 GB/s\\
Power&45 W&204 W&81 W\\
\end{tabular}
\end{ruledtabular}
\end{table*}
Hadamard gate is represented by a 2 by 2 matrix and controlled phase operators are 4 by 4 matrices. A system of $n$-qubit is expressed by $2^n$ values with a tensor product. The QFT operator is also specified by a matrix of $2^n \times 2^n$. Unfortunately, after measurements, collapsed states cannot easily be converted to qubits. Therefore, it is easier to multiply $2^n \times 2^n$-QFT matrix by $2^n$ state vector.

In Figure~\ref{Fig2}(a) we give a block configuration and running order in GPU. Here, the number of blocks is equal to $2^n/(\text{block size})$. Each thread computes each line which requires $2^n$ complex multiplications as follows:
\begin{equation}\label{bb}
 V_k=\sum^{q}_{j=0}e^{i2 \pi jk/q} V_j.
\end{equation}

\begin{table}[b]
\caption{Theoretical data transfering time for our GPUs. Here, \# of bytes $= (2\times(2^{2n}\times2^{2n})+(4\times2^{2n}))\times4$.}\label{table2}
\begin{ruledtabular}
\begin{tabular}{rrrr}
\multicolumn{1}{c}{\textbf{n}}&\multicolumn{1}{c}{\textbf{\# of bytes}}&\multicolumn{1}{c}{$T_{285}$[s]}&\multicolumn{1}{c}{$T_{970m}$[s]}\\ \hline

7&	2147745792&	     0.012580152&	  0.016668701\\

8&	34360786944&	 0.201264004&   	  0.266674805\\

9&	5.4976$\times 10^{11}$&	     3.220150354&	  4.266699219\\

10&	8.7961$\times 10^{12}$&	51.52211085&	  68.26679688\\
\end{tabular}
\end{ruledtabular}
\end{table}
When we create more blocks to decrease computation load on the blocks, we need to allocate more space to handle semi-results, as shown in Figure~\ref{Fig2}(b).

Another idea is to compute complex QFT matrix before the transform operation. In this case only two complex matrix and two complex state vector will be transferred between CPU and GPU. Considering the memory bandwidth of GTX 285 and GTX 970m GPUs in Table~\ref{table1}, we can conclude that wheter our kernel has memory bottleneck. According to Table~\ref{table2}, GPU transferring time is reasonable. But for factoring  large integers the memory capacity is not efficient. For a 8-qubit system about 32 GB space required which is unfeasible with current devices.

\section{\label{sec:results} Experimental Results }
In this section, we describe the experimental environment and discuss the performance of our implementations.
We have utilized GPUs and CPU for benchmark tests. The differences in micro-architectural philosophy between NVIDIA GPUs and CPU used for all test results are shown in Table~\ref{table1}. The experimental computer system contains 8 GB of memory whereas the GPU device has only 3 GB. This difference between the memory sizes of host and the GPU requires frequent data transfers.
\begin{table*}[!t]
\caption{The performance results for factorization. Here, all the execution times are in seconds. $T_{H,FH,Liquid}$ represent to the timing result of Hayward, Fast-Hayward and Liquid, respectively. $T_{G285,G970m}$ refer to the timing result of GPU GTX285 and GTX970m, respectively.}
\centering
\label{table3}
\begin{ruledtabular}
\begin{tabular}{lcrrrrrrr}
 &~~~~~~& \multicolumn{1}{c}{~~\textbf{n}~} & \multicolumn{1}{c}{~~~\textbf{Cofactors}} & \multicolumn{1}{c}{\textbf{$T_H$}} & \multicolumn{1}{c}{\textbf{$T_{FH}$}} & \multicolumn{1}{c}{\textbf{$T_{Liquid}$}} & \multicolumn{1}{c}{\textbf{$T_{G285}$}} & \multicolumn{1}{c}{\textbf{$T_{G970m}$}} \\\hline
 &~& 77 & 7$\times$11 & 111.462 & \cellcolor[HTML]{96FFFB}3.205 & \cellcolor[HTML]{96FFFB}47.125 & \cellcolor[HTML]{96FFFB}0.725 & \cellcolor[HTML]{96FFFB}1.167 \\
 && 143 & 11$\times$13 & 114.015 & \cellcolor[HTML]{96FFFB}34.481 & \cellcolor[HTML]{96FFFB}189.523 & \cellcolor[HTML]{96FFFB}3.236 & \cellcolor[HTML]{96FFFB}1.791 \\
 && 323 & 17$\times$19 & 1915.227 & \cellcolor[HTML]{96FFFB}462.850 & \cellcolor[HTML]{96FFFB}1171.650 & \cellcolor[HTML]{96FFFB}45.424 & \cellcolor[HTML]{96FFFB}21.375 \\
 && 551 & 19$\times$29 & \cellcolor[HTML]{999999} & \cellcolor[HTML]{999999} & \cellcolor[HTML]{999999} & 716.100 & \cellcolor[HTML]{999999} \\
&\multirow{-5}{*}{\textbf{\rotatebox[origin=c]{90}{ \parbox[c]{1.9cm}{\centering 2 cofactors}}}} & 589 & 19$\times$31 & \cellcolor[HTML]{999999} & \cellcolor[HTML]{999999} & \cellcolor[HTML]{999999} & 952.166 & \cellcolor[HTML]{999999} \\\hline
 && 231 & 3$\times$7$\times$11 & 459.258 & \cellcolor[HTML]{96FFFB}60.218 & \cellcolor[HTML]{96FFFB}214.320 & \cellcolor[HTML]{96FFFB}11.857 & \cellcolor[HTML]{96FFFB}5.725 \\
 && 255 & 3$\times$5$\times$17 & ~~1812.330 & \cellcolor[HTML]{96FFFB}115.955 & \cellcolor[HTML]{96FFFB}195.579 & \cellcolor[HTML]{96FFFB}11.568 & \cellcolor[HTML]{96FFFB}5.833 \\
 && 399 & 3$\times$7$\times$19 & \cellcolor[HTML]{999999} & \cellcolor[HTML]{96FFFB}4846.131 & \cellcolor[HTML]{96FFFB}1126.509 & \cellcolor[HTML]{96FFFB}180.153 & \cellcolor[HTML]{96FFFB}83.675 \\
 && 423 & 3$\times$3$\times$47 & \cellcolor[HTML]{999999} & \cellcolor[HTML]{96FFFB}5130.147 & \cellcolor[HTML]{96FFFB}1179.300 & \cellcolor[HTML]{96FFFB}180.485 & \cellcolor[HTML]{96FFFB}~~~~85.140 \\
&\multirow{-5}{*}{\textbf{\rotatebox[origin=c]{90}{ \parbox[c]{1.9cm}{\centering 3 cofactors}}}} & ~~539 & 7$\times$7$\times$11 & \cellcolor[HTML]{999999} & 11645.820 & ~6705.252 &~~714.752 & \cellcolor[HTML]{999999} \\\cline{1-9}
&\multicolumn{4}{c}{\textbf{Speed-up}} & \textbf{52.5} & \textbf{20.5} & \textbf{2.1} & \textbf{1.0}\\
\end{tabular}
\end{ruledtabular}
\end{table*}
Theoretical performance of a computing device could be estimated by multiplying the number of cores, core clock speed, FMA and SIMD. The theoretical values of our CPU and GTX970m are 27.66 Gflop (Giga Floating-Point Operation) and 2657 Gflop per second , respectively.

We have tested our GPU implementation to find the prime factors of 10 integers. We compare the timing results with Hayward, optimized-Hayward~\cite{Hayward} and Liquid~\cite{liquid}. Hayward and optimized-Hayward implementations run on single CPU-core. Liquid is the Microsoft quantum simulation platform and it is fairly optimized for CPUs. Table~\ref{table3} shows the performance results from factorization, using Intel I7-2760QM, NVIDIA GTX285 and NVIDIA 970m. In this table, gray cells indicate that no results are obtained due to a memory fault occurred or the computation lasted over 3 hours. Furthermore, speedup-values are calculated according to run-time values in blue cells and timing values of GTX970m are determined as reference. As seen in Table~\ref{table3}, GTX970m has $52.5\times$ speedup over Fast-Hayward, $20.5\times$ speedup over Liquid and $2.1\times$ speedup over GTX285. These results show a significant performance improvement when using a GPU. Consequently, it is clear that the algorithm is achieving its goal of accelerating of the factorization computation on the GPU.

\section{\label{sec:conc} Conclusion}
We have proposed a GPU implementation of Quantum Fourier Transform. For this we have presented a simulator for Shor's quantum algorithm on GPUs and compare our results with Liquid and two CPU-implementations.
Due to transferring data is not a bottleneck (see Table~\ref{table2}), our QFT kernel is limited by  compute-bound of the GPUs. We achieved 52.5x, 20.5x speedup against the classical transform function and Liquid respectively.

\end{document}